# Superconductivity in Semimetallic $Bi_3O_2S_3$

L. Li[1], D. Parker[1], P. Babkevich[2], L. Yang[2, 3], H. M. Ronnow[2], and A. S. Sefat[1]

[1] Materials Science & Technology Division, Oak Ridge National Laboratory, Oak Ridge, Tennessee 37831, USA

[2] Laboratory for Quantum Magnetism (LQM), École Polytechnique Fédérale de Lausanne (EPFL), CH-1015 Lausanne, Switzerland

[3] Laboratory of Physics of Complex Matter, École Polytechnique Fédérale de Lausanne (EPFL), CH-1015 Lausanne, Switzerland

## Abstract

Here we report further investigation on the thermodynamic, and transport properties, and the first assessment of theoretical calculations for the $BiS_2$-layered $Bi_3O_2S_3$ superconductor. The polycrystalline sample is synthesized with the superconducting transition temperature of $T_c^{onset}$ = 5.75 K and $T_c^{zero}$ = 4.03 K ($\approx T_c^{mag}$), that drops to 3.3 K by applying hydrostatic pressure of 6 kbar. Density-of-states calculations give substantial hybridization between Bi, O and S, with the largest Bi component at DOS, which supports the idea that $BiS_2$ layer is relevant for producing electron-phonon coupling. The electron charge carrier concentration of only 1.5 x $10^{19}$ $cm^{-3}$ for $Bi_3O_2S_3$ is additionally suggestive of strong electron-phonon interaction in Bi-O-S system. The analysis of Seebeck coefficient results strongly suggests that $Bi_3O_2S_3$ is a semimetal. In fact, the semimetallic or narrow band gap behavior may be general in the $BiS_2$-layered class of materials, including that of $Bi_4O_4S_3$.

# 1 Introduction

Understanding reasons for superconductivity in Bi-O-S system is important, especially since high-temperature superconductivity is seen in other Bi-based materials such as $Ba_{1-x}K_xBiO_3$ and $Bi_2Sr_2Ca_2Cu_3O_{10+x}$ [1]. The superconducting members of Bi-O-S system are based on $BiS_2$-layers with large unit cells composed of various spacer layers. The members include $Bi_4O_4S_3$ with $T_c$ = 4.4 K, $H_{c2(0)}$ ~ 3 T, and dominant-electron concentration of ~ 2 × $10^{19}$ $cm^{-3}$ at 300 K [2-4]. The enhanced density of states at Fermi level from Bi $6p_x$ and $6p_y$, and the strong Fermi surface nesting of quasi-one-dimensional bands are predicted to play a role in Cooper pairing [2]. Theoretical tight-binding calculations on a similar $BiS_2$-layer-based system of $LaOBiS_2$ (F-doping; $T_c$ = 10.6 K) [5] shows that dominating bands for electron conduction and superconductivity are those derived from four Bi $6p$-S $3p$ orbitals that give nearly half-filled, electron-hole symmetric, quasi-one-dimensional bands [6] that give nesting of the Fermi surface. Other members of Bi-O-S system include $Bi_6O_8S_5$ [7], and $Bi_3O_2S_3$ [8, 9]. Most recently, critical temperature at $T_c$ = 4.5 K was discovered in $Bi_3O_2S_3$ [8] that appears to be an unconventional superconductor with a low $T_c^{onset}$ = 5.8 K, upper critical field of $H_{c2}$ = 4.8 T and low electron-dominant carriers of ~ $10^{19}$ $cm^{-3}$ [9]. This manuscript focuses on $Bi_3O_2S_3$ by advancing investigations into reasons of superconductivity through theoretical calculations and bulk thermodynamic and transport property results. A schematic image of the unit cell is shown in the right inset of Figure 1a, depicting layers of $(BiS_2)_2$ (rock salt-type), $Bi_2O_2$ (fluorite-type) and $S_2$. We find that $T_c$ decreases by application of 6 kbar, similar to $Bi_4O_4S_3$ [10]. From Seebeck coefficient results and calculations, we find that $Bi_3O_2S_3$ is a semimetal and the band overlap is most likely in the range of 27 – 41 meV. The unusual low carrier concentration of only 1.5 x $10^{19}$ $cm^{-3}$ obtained from Hall measurements suggests a particularly strong interaction and one that could presumably be strengthened significantly via charge doping, if feasible, leading to larger $T_c$ for this new Bi-S-O system.

# 2 Experimental

Polycrystalline sample of $Bi_3O_2S_3$ was prepared by a conventional solid state reaction method according to the literature precedent [8, 9]. The $Bi_2O_3$ powders and S pieces were weighed by targeting "$Bi_{12}O_{18}S_{13}$", fully mixed, ground and then pressed into pellets. The pellets were sealed in an evacuated silica tube and placed in a pre-warmed furnace at 430 °C. After four hours, the sample was quenched in ice water. The sample was removed from the tube, reground and repressed, and then sealed in another quartz tube. The tube was placed back in the pre-heated furnace at 430 °C. After 14 hours, the sample was quenched down to 0 °C with ice water. After another regrinding and pelletizing, the pellets was sealed in an evacuated silica tube and placed in a pre-heated furnace at 520 °C for 17 hours, and then quenched again. The resultant sample looked black and dense. The x-ray diffraction (XRD) pattern was collected by a PANalytical X'Pert PRO MPD x-ray powder diffractometer with Cu Kα radiation. The Rietveld refinement of the XRD patterns was done with FullProf software suite, using the reported Wyckoff positions [8] and $I4/mmm$ space group. The electrical and thermal transport measurements were performed in a Quantum Design (QD) Physical Property Measurement System (PPMS), using a standard four-probe method. Magnetization measurements were performed in a QD Magnetic Property Measurement System (MPMS). The application of pressure was performed in a Mcell 10 pressure cell (easyLab Technologies) inside the MPMS; the applied pressure was estimated from the $T_c$ of the Pb manometer. First principles calculations were performed using the linearized augmented plane-wave code WIEN2K [11], employing the generalized gradient approximation (GGA) of Perdew, Burke and Ernzerhof [12]. $RK_{max}$ was set to 9.0, where R is the smallest muffin-tin radius and $K_{max}$ the largest planewave wavevector. Muffin tin radii for all calculations, in units of the Bohr radius $a_0$, were as follows: Bi – 2.31, O – 1.89, S – 2.08 except for the dimerized S, which were taken as 1.8. All internal coordinates were optimized until forces were less than 2 mRyd/ Bohr, while lattice parameters were taken from experiment. 1000 or more k-points in the full Brillouin zone were used for all calculations. Spin-orbit coupling was included for all calculations excepting the optimization. We also performed Boltzmann transport calculations of the thermopower using the Boltztrap code [13]. For these calculations as many as 40,000 k-points in the full Brillouin zone were used.

## 3 Results and discussion

Figure 1a shows the XRD pattern for polycrystalline $Bi_3O_2S_3$, which looks very similar to that reported in Ref [8]. Only tiny impurity amounts of $Bi_2OS_2$ [8], $Bi_2S_3$ and Bi are present, which have also been taken into account for a refinement that improved the fitting. The refined lattice constants are $a$ = 3.9645(5) Å and $c$ = 41.3165(4) Å. Impurity phases were estimated using the Hill and Howard method [14], and their total contents were less than 8% by mass.

Temperature dependence of resistivity is shown in Fig. 1b. In zero field, $T_c^{onset}$ = 5.75 K and $T_c^{zero}$ = 4.03 K. Upon application of 1 T magnetic field it can be seen that the $T_c^{onset}$ shift to lower temperature and the sample does not reach zero resistivity down to 2 K. Inset of Fig. 1b shows resistivity up to room temperature. It displays a typical metallic behavior, similar to that of $Bi_4O_4S_3$[2]. Temperature dependence of magnetic susceptibility is shown in Fig. 1c. The sharp transition in $\chi$(T) indicates that the sample is rather homogeneous. The value of $T_c$, defined by the divergence of zero-field-cooled and field-cooled results, is about 4.1 K. The shielding volume fraction at 1.8 K is ~ 93%. Higher field measurements (inset of Fig. 1c) reveals that $Bi_3O_2S_3$ exhibits a diamagnetic behavior without any abrupt drop or anomaly, indicating there is no magnetic ordering in this system. Compressing the structural lattice parameters by applying external pressure is believed to be an effective method of exploring the optimized condition for superconductivity. To this end, we have performed a pressure study of the temperature dependence of susceptibility for $Bi_3O_2S_3$ below 5 K. The hydrostatic pressure does not improve the superconductivity of the sample, while on the contrary, the transition temperature $T_c^{mag}$ ~ 4 K at ambient pressure decreases to 3.3 K at 6 kbar, as shown in Fig. 1d. Although similar $T_c$ suppression was seen in $Bi_4O_4S_3$ [10], further theoretical work on DOS and structural analysis at high-pressure could be useful for a better understanding of this effect.

The band structure calculation of $Bi_3O_2S_3$ is shown in Fig. 2a; it suggests that this material is a semiconductor with a very small band gap (approximately 25 meV) and substantial band degeneracy, particularly in the conduction bands, where a second band minimum occurs less than 50 meV above the primary band minimum, with all these extrema at or very near the N point. The calculated density-of-states for this material is shown in Fig. 2b. The plot shows a comparatively heavy band just below the valence band maximum, and a somewhat lighter band in the conduction band. Given the Hall measurements indicate *n*-type behavior [9] (Fig. 1b inset), we focus our attention on the conduction band. This shows substantial hydridization between the Bi, O and S. Previous papers have argued for the importance of the $BiS_2$ layers in generating electron-phonon coupling [2], and this hybridization is supportive of this idea.

One natural interpretation of our experimental data in light of these plots is that $Bi_3O_2S_3$ is indeed a narrow band semiconductor, with an extrinsic n-type doping level of approximately 1.5 x $10^{19}$ $cm^{-3}$, as suggested by the Hall data. However, one immediately finds two problems with this explanation. Firstly, this carrier concentration is the same as found in the prior work (Ref. [9]), and were this simply a semiconductor such a coincidence in extrinsic doping levels would seem rather unlikely. Secondly, and more compellingly, the thermopower data presented in Fig. 3 shows strong evidence of semi-metallic, rather than semiconducting behavior. In Fig. 3 we plot the experimental Seebeck data, along with two theoretical fits to the data. These fits were generated using the first principles calculated band structure, with the thermopower calculated via the Boltzmann transport code Boltztrap [13], using the constant scattering time approximation, which has shown considerable success in describing doped semiconductors.

The first fit (dashed line) was plotted using the first principles-calculated band gap of 30 meV for a chemical carrier concentration of $n = 1.2 \times 10^{19}$ cm$^{-3}$, comparatively near the Hall data generated value (note that these two need not be exactly equal). This fit shows a low temperature thermopower slope S/T (T < 50K) approximately *ten* times the measured value, and then rapidly enters the bipolar regime soon thereafter. Neither of these behaviors resembles the gently increasing curve in the experimental data (we suspect this data would show a maximum at T ~ 350 K). We have also tried fitting the data with a much heavier chemical doping ($n = 1.5 \times 10^{20}$ cm$^{-3}$, not shown). While one can fit the initial slope of S(T) in this way, the thermopower magnitude at temperatures above 50 K continues to increase linearly, reaching magnitudes much larger than the experimental value. This is not surprising considering that $E_F$ (T=0) in this case is approximately 150 meV above the conduction band minimum, and nearly 180 meV above the valence band maximum, so that bipolar effects are minimal in this temperature range at this doping. The second curve (black solid line) was generated using the first principles calculated band structure with a "scissors shift" employed. In this "scissors shift" the conduction and valence band structures are taken as unchanged, but the band gap is adjusted. The plot indicates that band overlap of 0.003 Rydberg=41 meV (i.e. a -41 meV band gap) produces a good fit to the experimental data. We take this as strong evidence that $Bi_3O_2S_3$ is in fact a semimetal, not a semiconductor. Note that we are also able to fit the data with a somewhat smaller band overlap of 27 meV, so we can say that the actual band overlap is most likely in the range of 27 – 41 meV.

With the electronic structure established, let us consider the superconductivity itself in more detail. We note that superconductivity with $T_c$ of ~ 5 K is rather unusual in a material with a carrier concentration of only $1.5 \times 10^{19}$ cm$^{-3}$, and is suggestive of strong electron-phonon interaction. Let us now make an approximate estimate of this interaction, using present and previous data. In Ref. [9] a value of the T-linear specific heat constant $\gamma$= 1.65 mJ/mol-K$^2$ was reported. On a density-of-states basis this is 0.7/eV-formula unit, or 0.23/eV-BiS unit, noting that previous references [6] have argued the BiS$_2$ layers to be the relevant component producing electron-phonon coupling, the presumed pairing source. Assuming a modified BCS expression $T_c \sim \omega_D \exp(-(1+\lambda)/\lambda)$, including the Eliashberg mass renormalization and neglecting the Coulomb pseudopotential $\mu^*$, a 5 K $T_c$ with the 188 K Debye temperature estimated in Ref. [9] corresponds to an electron-phonon coupling constant $\lambda = N_oV$ of 0.38. This yields a modified electron-phonon interaction potential V of 1.5 eV per BiS unit. This rough calculation suggests a particularly strong interaction and one that could presumably be strengthened significantly via charge doping, if feasible, leading to larger $T_c$.

We note that the same features suggestive of semimetallic behavior in $Bi_3O_2S_3$ – the increasing thermal conductivity above the Debye temperature and likely saturation and turning-over of the temperature dependent thermopower – appear to be present in the related material $Bi_4O_4S_3$ [15]. In this case we cannot easily distinguish between a semimetal and narrow band gap semiconductor, but note that both cases are strongly at odds with the band structure calculations for this material published in Ref. 2, which depict a band edge some 0.6 eV above $E_F$ – i.e. metallic behavior. Hence the observance of semimetallic or narrow band gap behavior may be more common or general in this class of materials than previously thought.

# 4 Conclusions

In summary, we have synthesized and confirmed the bulk superconductivity by both resistivity and magnetic susceptibility measurements in the new $BiS_2$-layer-based superconductor $Bi_3O_2S_3$.. Hall results suggest that electronic charge carriers are dominant in normal state with the unusual low concentration of only 1.5 x $10^{19}$ $cm^{-3}$, indicating the sample has particularly strong electron-phonon interaction, as also confirmed by DOS calculations. The thermopower data shows strong evidence of semi-metallicity for this new Bi-S-O system, and the band overlap is most likely in the range of 27 – 41 meV.

# Acknowledgement

This work was primarily supported by the U. S. Department of Energy, Office of Science, Basic Energy Sciences, Materials Science and Engineering Division, and partially by ORNL's LDRD funding (all theoretical modeling and calculations). The pressure measurements were funded by the Swiss National Science Foundation its Sinergia Network Mott Physics Beyong the Heisenberg Model and the European Research Council project CONQUEST; in addition, Jonathan White is thanked for his assistance.

## Figures & Captions:

**Figure 1.** (a) Powder X-ray diffraction pattern for $Bi_3O_2S_3$. Red circles represent observed data; black and green solid lines represent the calculated intensity and difference between the observed and calculated intensity; blue vertical bars indicate the Bragg reflection positions for $Bi_3O_2S_3$. * and # indicate the peaks of $Bi_2S_3$ and Bi, respectably. Inset shows the schematic image of the crystal structure of $Bi_3O_2S_3$.

(b) Temperature dependence of resistivity for $Bi_3O_2S_3$ measured below 10 K, at 0 T and 1 T applied magnetic fields. Inset is the temperature dependence of resistivity and charge carrier concentration for $Bi_3O_2S_3$ up to room temperature at 0 T.

(c) Temperature dependence of magnetic susceptibility for $Bi_3O_2S_3$ measured below 10 K, at H = 5 Oe applied magnetic fields. The inset displays the temperature dependence of magnetic susceptibility for $Bi_3O_2S_3$ up to room temperature obtained at 1 T applied magnetic fields.

(d) Temperature dependence of magnetic susceptibility for $Bi_3O_2S_3$ measured below 4.5 K, with applying hydrostatic pressure of 0 kbar (red circle) and 6 kbar (blue triangle), respectively.

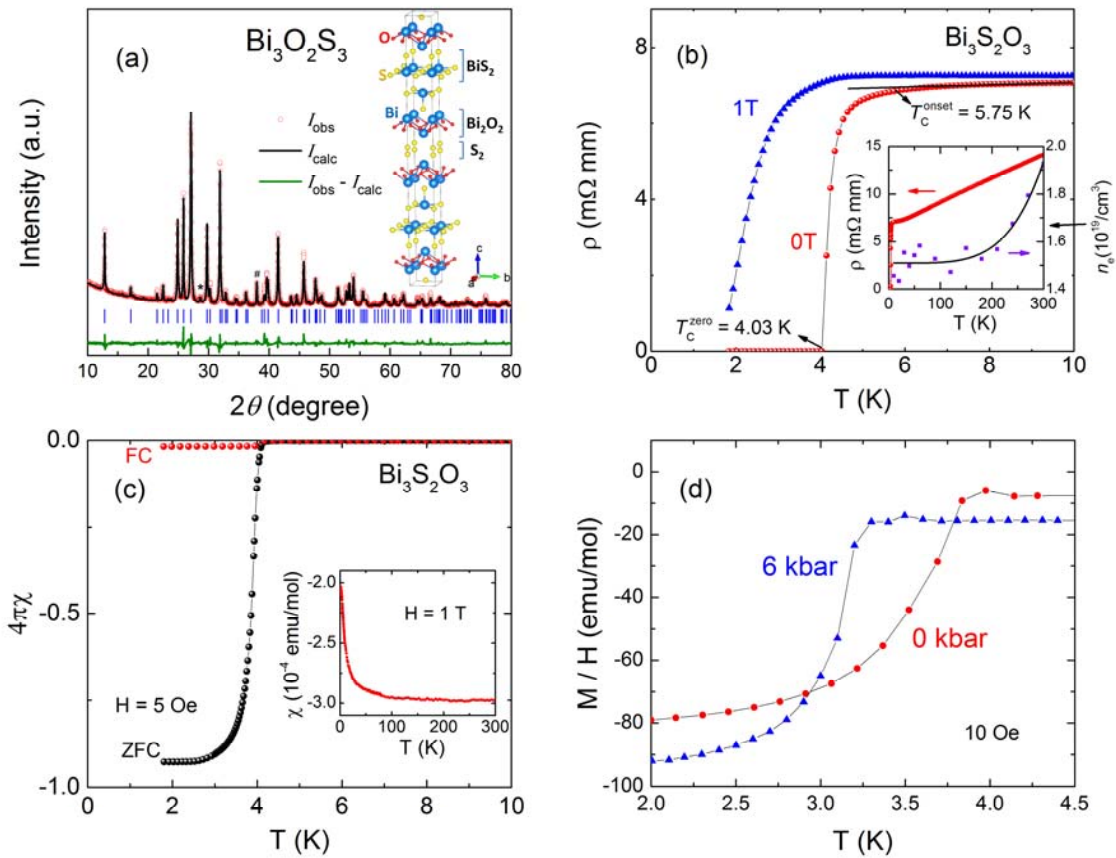

**Figure 2.** (a) The calculated band structure of $Bi_3O_2S_3$, within the body-centered tetragonal Brillouin zone. The energy zero is set to the valence band maximum. Note the very small band gap - approximately 25 meV - at the N-point. (b) The calculated density-of-states of $Bi_3O_2S_3$. The energy 0 is set to the valence band maximum.

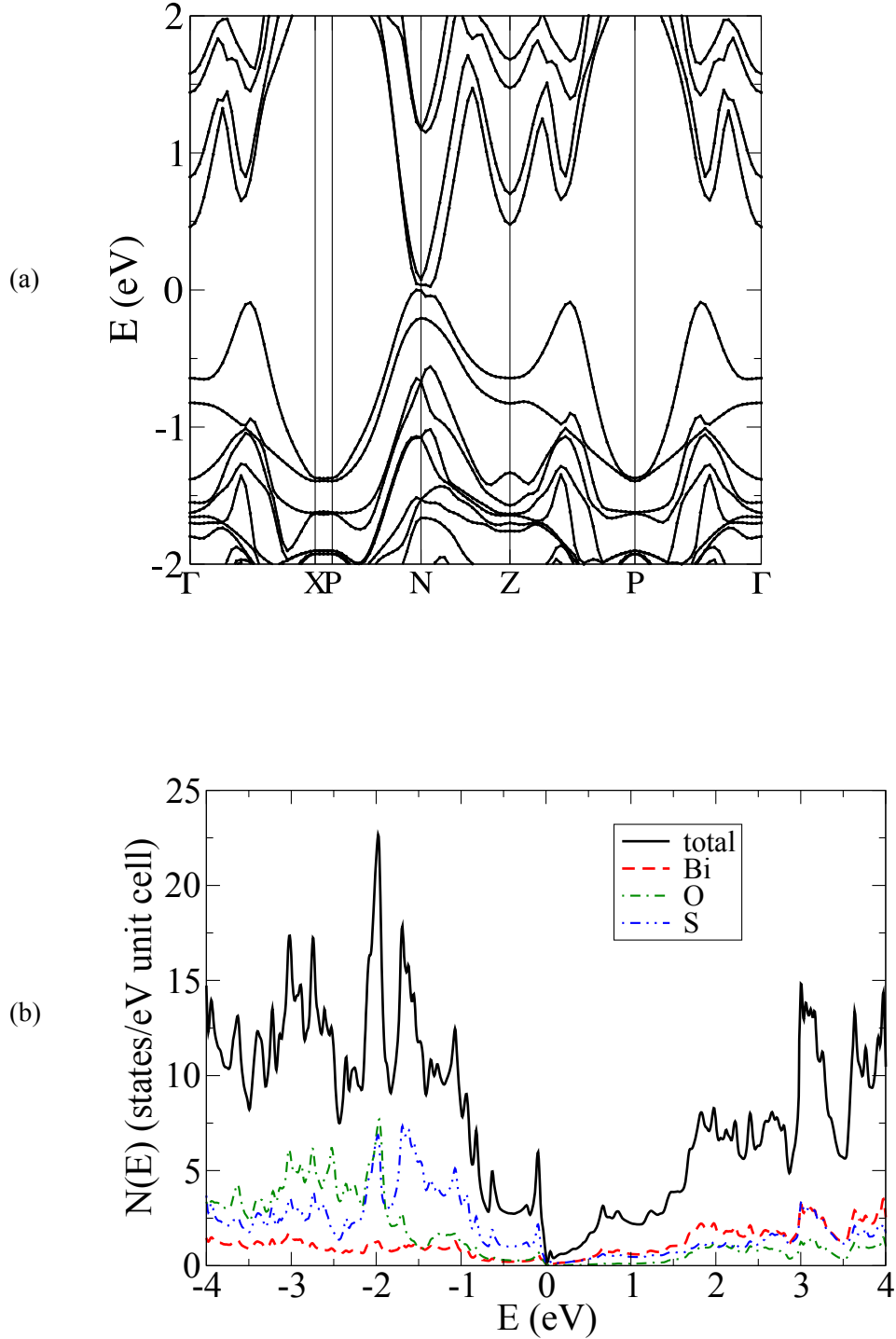

**Figure 3.** The measured thermopower (diamonds) and calculated thermopower under two scenarios: dashed line, the semiconducting band structure; solid line, an assumed semimetallic band structure with a band overlap of 41 meV. Inset shows the temperature dependence of the Seebeck coefficient and thermal conductivity up to room temperature.

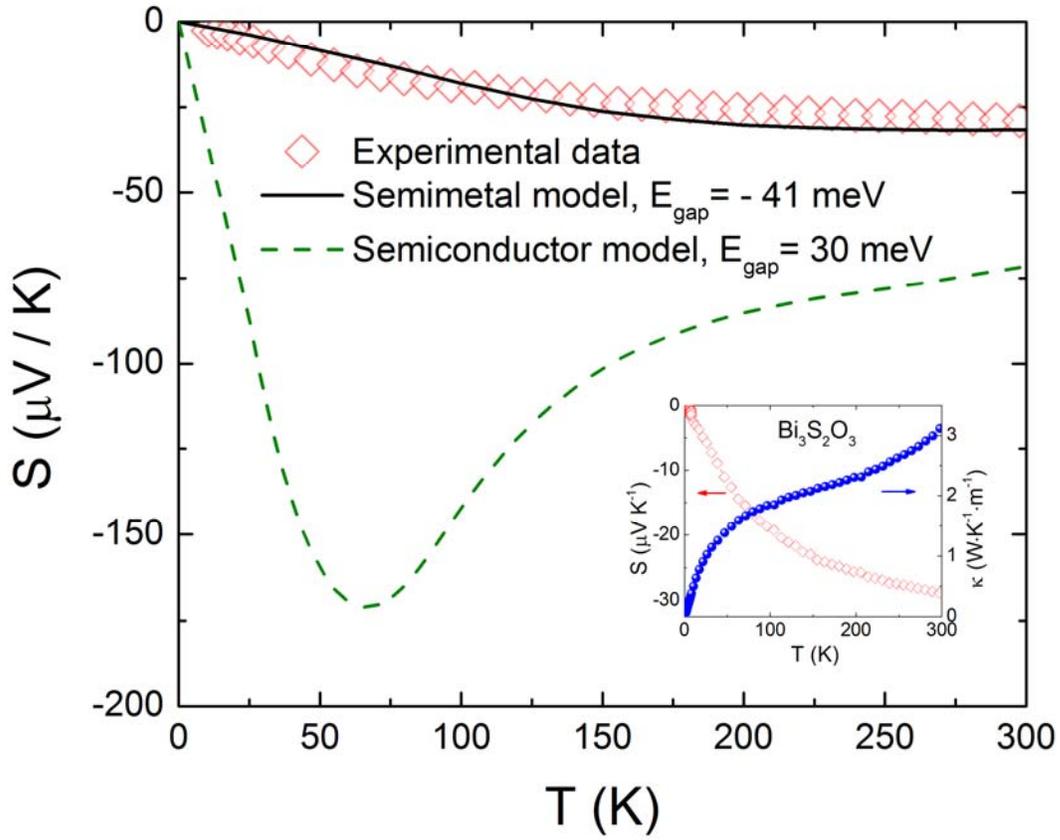